\documentclass[aps,twocolumn, floats,preprintnumbers,showpacs,prl]{revtex4}
\usepackage{graphicx}
\usepackage{color}
\usepackage{soul}
\usepackage{latexsym}
\usepackage{amsmath}
\usepackage{amsfonts}
\usepackage{amssymb}
\def\beq{\begin{equation}}
\def\eeq{\end{equation}}
\def\bea{\begin{eqnarray}}
\def\eea{\end{eqnarray}}

\def\lsim{\mathrel{\raise.3ex\hbox{$<$\kern-.75em\lower1ex\hbox{$\sim$}}}}
\def\gsim{\mathrel{\raise.3ex\hbox{$>$\kern-.75em\lower1ex\hbox{$\sim$}}}}
\def\ifmath#1{\relax\ifmmode #1\else $#1$\fi}

\def\gev{~{\mbox{GeV}}}

    \def\fillboxx#1#2{\hbox to #1{\vbox to #2{\vfil}\hfil}   }

\def\gev{~{\rm GeV}}

\def\cnone{\wt\chi^0_1}
\def\cndark{\wt\chi_{\rm dark}}

\def\wt{\widetilde}

\def\dark{{\rm dark}}

\newcommand{ \slashchar }[1]{\setbox0=\hbox{$#1$}   % set a box for #1
   \dimen0=\wd0                                     % and get its size
   \setbox1=\hbox{/} \dimen1=\wd1                   % get size of /
   \ifdim\dimen0>\dimen1                            % #1 is bigger
      \rlap{\hbox to \dimen0{\hfil/\hfil}}          % so center / in box
      #1                                            % and print #1
   \else                                            % / is bigger
      \rlap{\hbox to \dimen1{\hfil$#1$\hfil}}       % so center #1
      /                                             % and print /
   \fi}     

\begin{document}
\preprint{FERMILAB-PUB-09-026-T}

\title{Measuring the Dark Force at the LHC 
}

\author{Yang Bai${}^{a}$ \thanks{email: bai@fnal.gov}}
\author{Zhenyu Han${}^{b}$ \thanks{email:zhenyuhan@physics.ucdavis.edu}}
\address{  \vspace{3mm}
${}^{a}$Theoretical Physics Department, Fermilab, Batavia, Illinois 60510,
${}^{b}$Department of Physics, University of California, Davis, CA 95616}

%\date{\today}


\begin{abstract}
A long-range ``dark force'' has recently been  proposed  to mediate the
dark matter (DM) annihilation. If DM particles are copiously
produced at the Large Hadron Collider (LHC), the light dark force mediator
will also be produced through radiation. We demonstrate
how and how precise we can utilize this fact to measure the coupling
constant of the dark force. The light mediator's mass is measured
from the ``lepton jet'' it decays to. In addition, the mass of the
DM particle is determined using the  $m_{T2}$
technique. Knowing these quantities is critical for calculating the
DM relic density.    

\end{abstract}
\maketitle
%\keywords{}
{\bf Introduction.}
Recently, there have been strong indications of indirect detections of
dark matter (DM) particles. In particular, the payload for antimatter matter exploration  and light-nuclei astrophysics (PAMELA) 
experiment has observed a sharp turn-over of positron fraction in the
cosmic rays in the 10--100~GeV range~\cite{Adriani:2008zr}, which can
be naturally explained by DM-DM annihilation to  
electron-positron pairs with a large ``boost factor''. 
%This observation is supported by the ATIC~\cite{ATIC} and
%PPB-BETS~\cite{Torii:2008xu} balloon experiments that have also seen
%excesses in the electron/positron spectra~\cite{FermiLAT}. 
In an interesting DM scenario \cite{ArkaniHamed:2008qn},  the boost factor is obtained 
through the Sommerfeld enhancement effect \cite{Cirelli:2008pk}, the
consequence of a long-range attractive force between DM particles. A
light mediator $a_{\rm dark}$  with an $O({\rm GeV})$ mass provides the attractive
force. Given its small mass, $a_{\rm dark}$ must couple to the
standard model (SM) very weakly (hence the force it mediates is given the
name ``dark force''), otherwise its effects should have long been
seen. Accordingly, its coupling to the DM has to be of
order unity, to obtain the required annihilation rate. 
The lack of anti-proton excess in the PAMELA
experiment~\cite{Adriani:2008zq} can be explained if $a_{\rm dark}$ is
so light that its decays to hadrons are kinematically
forbidden~\cite{ArkaniHamed:2008qn}, or if it dominantly couples to
the SM leptons~\cite{bai_fox}. 

The above DM scenario also has unusual signatures at the LHC
\cite{DMLHC}. If $a_{\rm dark}$ decays 
within a collider detector and  dominantly to leptons, we will
be able to see ``lepton jets'', where two or more leptons are boosted and 
collinear to one another. Once events with lepton jets as well as
large missing energy are observed, it is critical to test whether they
come from the same theory that explains the positron/electron
anomalies. The purpose of this letter is to discuss how to extract the
relevant information from 
the LHC measurements to test this new class of DM models. In
particular, we demonstrate how to measure the DM  particle mass, the
light mediator mass, and the coupling constant $g$ of the DM-$a_{\rm
  dark}$ interaction. Once these quantities are determined, we will be
able to calculate DM-DM annihilation rate. This allows us to compare
with the results of DM searches as well as to verify if we obtain the
correct DM relic density assuming it is thermally produced.  

\begin{figure}[ht!]
\centerline{ \hspace*{0cm}
\includegraphics[width=0.4\textwidth]{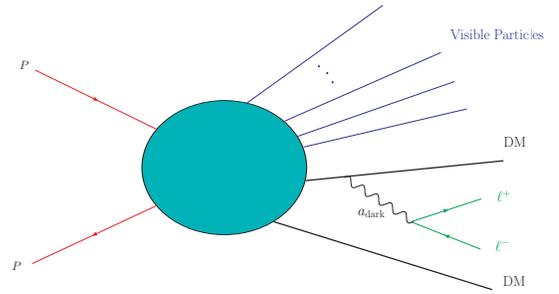}
} 
\caption{The schematic Feynman diagram of the dark matter radiating a
  light mediator $a_{\rm dark}$, which decays to two leptons.} 
\label{fig:DMrad}
\end{figure}
To be able to perform these measurements, it is crucial that the DM
particle can be produced at the LHC, which we assume to be the 
case. Once it is produced, it has a significant probability  to
radiate an extra particle $a_{\rm dark}$ (Fig.~\ref{fig:DMrad}),
because ${a_{\rm dark}}$ is light and the DM-$a_{\rm dark}$ coupling
is large. Therefore, for whatever process  
to produce the DM particle, there is a corresponding process
with an extra ${a_{\rm dark}}$ produced through final state radiation. The
production rate ratio of the two processes in general depends on
$M_{a_{\rm dark}}$, the coupling $g$ and the DM mass
$m_{\rm DM}$, which can be written as  
\bea
&&\frac{\sigma (p\,p\,\rightarrow\, X\,{\rm DM}\, {\rm
    DM}\,a_{\rm dark})}{\sigma(p\,p\,\rightarrow\,X\,{\rm DM}\,{\rm DM})} 
\nonumber \\&&\hspace{1.5cm}
\approx
 C\, \frac{g^{2}}{4\,\pi^{2}}\,\log{\left(
  \frac{q^{2}}{M_{a_{\rm dark}}^{2}}  \right)}\,\log{\left(
  \frac{q^{2}}{m_{\rm DM}^{2}}  \right)}\,. 
\label{eq:ratio}
\eea
Here $X$ represents visible particles; $C$ is a process-dependent
coefficient. This $q^{2}$ dependent result is known as the Sudakov
double logarithm. If other particles charged under the dark force are
produced, they can also radiate an $a_{\rm dark}$ in which case 
$m_{\rm DM}$ in Eq.~(\ref{eq:ratio}) should be replaced by the relevant
mass. As we will show shortly,  $M_{a_{\rm dark}}$ and $m_{\rm DM}$
can be determined independently, which enables us to
further extract the coupling.

To be concrete, in this letter we envisage that the LHC signal events
are produced in a two-step process. The first step is entirely
analogous to ordinary models with missing particles, such as the Minimal
Supersymmetric Standard Model (MSSM) with R-parity or Universal Extra
Dimension with KK-parity. We will adopt the nomenclature of the MSSM
although a general setup is understood. Large cross sections are
achieved when colored particles are produced and subsequently decay to
the lightest supersymmetric particle (LSP) in the MSSM. Due to the
small coupling between the DM and the SM, these decays are not
affected. In 
addition, we assume the MSSM LSP is heavier than the DM, which is
R-parity odd. Therefore, in the second step, the MSSM LSP 
decays to the DM plus some extra visible particles. In the following
we exemplify our method by analyzing a simple model in the above
scenario. After giving the essential ingredients of the
model, we discuss in turn the measurements of the mediator mass,
the DM mass and the coupling. We conclude the letter with a few
discussions.

{\bf The measurments.} The visible sector in our model is the
ordinary MSSM with a Bino-like neutralino LSP, $\cnone$. The dark
sector contains a $U(1)_{\dark}$ gauge group, and two Higgs
superfields with opposite charges under $U(1)_{\dark}$. The dark
sector interacts with the 
ordinary MSSM sector through a small kinetic mixing between the
$U(1)_{\rm dark}$ gauge superfield and the $U(1)_{Y}$ gauge
superfield. We assume that the dark sector LSP is the dark Higgsino,
$\cndark$, which is our DM candidate. Correspondingly,  the
dark gauge boson, which is identified as the mediator $a_{\rm dark}$,
provides an attractive force between two $\cndark$'s. $U(1)_{\rm 
  dark}$ is broken by the vacuum expectation values of the dark
Higgs fields (the lightest physical Higgs is denoted as $h_\dark$), which provide
$a_\dark$ a mass of $O(1\gev)$. Due to the kinetic
mixing, $a_\dark$ decays to two SM leptons, which are collinear with each
other and form a ``dark gauge boson jet'' (or $a$-jet). For the dark Higgs
$h_{\rm dark}$, we assume it has a mass $M_{h_{\dark}}\gtrsim 2
M_{a_{\dark}}$.  Therefore it mainly decays to two $a_\dark$'s which
subsequently decay to four leptons forming a ``dark Higgs jet'' (or
$h$-jet). Given the assumption that $\cndark$ is lighter than
$\cnone$, $\cnone$ mainly decays to $\cndark$ plus $h_{\dark}$ or
$a_\dark$ 
%\footnote{For simplicity we 
%assume the masses of the other Higgs fields are large so that the
%decays to them are suppressed. This assumption, together
%with the fact that the DM (Higgsino) is much heavier than
%$h_\dark$, indicates that this model is 
%fine-tuned. More complicated model building is
%needed to avoid this problem.}. 
$\cnone$ can also undergo a
three-body decay to $\cndark+h_{\dark}+a_\dark$. The ratio of the
three-body and the two-body decay widths can be measured and used to
determine the coupling $g$. For this purpose, it is enough to count
the number of events containing two $h$-jets and the number of events
containing two $h$-jets plus one $a$-jet.

% (Fig.~\ref{fig:SDMrad}).   
%\begin{figure}[ht!]
%\centerline{ \hspace*{0.2cm}
%\includegraphics[width=0.4\textwidth]{Feyn3.eps}
%} 
%\caption{An example Feynman diagram for events containing 10 final
 % state leptons from 2 $h_\dark$'s and 1 $a_\dark$.} 
%\label{fig:SDMrad}
%\end{figure}

Given the above setup, we choose the dark matter mass to be 600
GeV, consistent with the ATIC results. We also fix $M_{a_\dark}=1\gev$ and 
$M_{h_\dark}=3\gev$. The coupling constant $g$ is 0.40
to provide the correct DM relic density, which determines the boost factor
to be $O(100)$. In the MSSM sector, we choose the masses of the
gluino, squarks (restricted to the first two generations) and the MSSM
LSP, $\cnone$, to be 1200 GeV, 1000 GeV and 700 GeV, 
respectively. The spectrum is chosen such that the gluino directly
decays to quark plus squark and the squark only  directly decays 
 to quark plus $\cnone$. We generate the parton level events
in the squark/gluino pair
production channels with Madgraph/Madevents \cite{madgraph} for the
LHC at 14~TeV. The total cross section is 0.84~pb. The 
2-body and 3-body 
decays for $\cnone$ are performed with Calchep \cite{calchep} and all other
particles, including  $a_\dark$, $h_\dark$ and the other super
particles, are decayed with BRIDGE \cite{bridge}. Here, we assume that 
$a_{\dark}$ 100\% decays to two muons, and we will comment on the case
with $a_{\dark}$ decaying to electrons later. The parton level 
events are further processed with PYTHIA \cite{pythia} for
showering/hadronization, 
and PGS \cite{pgs} for detector simulation.       

The lepton jets are reconstructed as follows: all muons are first
sorted according to their $p_T$'s. Then we choose the 
muon with the highest $p_T$ in the list as a ``seed'' for the lepton jet,
and add muons within 0.2 $rad$ of the seed muon direction to the
lepton jet. Used muons are removed from the list. The ``seed
axis" is fixed in this procedure. We repeat this 
procedure until all muons are used. Lepton jets with 2 muons are
tagged as $a$-jets and lepton jets with 3 or 4 muons are tagged as
$h$-jets. Events containing untagged muons are discarded. An $H_T> 
500\gev$ cut is imposed to reduce the SM background, where $H_T$ is
defined as the scalar sum of all objects' $p_T$ (including
$\slashchar p_T$) in the events. We list the numbers of background and
signal events after cuts in Table~\ref{tab:cuts}. The SM backgrounds
coming mainly from $b\,\bar{b}$ and $t\,\bar{t}$ final states are
negligibly small after imposing a mass window cut on $a$-jets.
\begin{table}[htdp]
\renewcommand{\arraystretch}{1.6}
\begin{center}
\begin{tabular*}{0.48\textwidth} {@{\extracolsep{\fill}} c  c  c  c c  } 
\hline \hline
 & $b\,\bar{b}$ & $t\,\bar{t}$ & $W/Z$'s & Signal \\
 \hline
$H_T>500$~GeV &  1106\,k  &  1068\,k  &  17\,k      &  8303   \\
\hline
No. of muons $\geq$ 4 & 168    & 1890    & 13  &  7650 \\
\hline
No. of lepton jets $\geq$ 2&  70  &  36  & $< 1$  &  6523  \\
\hline
$|M_{a{\rm jet}}-1\gev|\leq 0.1$~MeV &  $1$  &  $1$  & $< 1$  &  5970  \\
\hline \hline
\end{tabular*}
\end{center}
\caption{The numbers of background and signal events after cuts. A
  10~fb$^{-1}$ luminosity and a 14~TeV center of mass energy are assumed for the LHC. The final signal events after cuts contain $5939$ two-body events and $31$ $2h1a$ three-body events.} 
\label{tab:cuts}
\end{table}%

{\it Measuring $M_{a_\dark}$ and $M_{h_\dark}$}. The mass measurements of the
particles $a_\dark$ and $h_\dark$ are straightforward since that they
are given respectively by the invariant masses of the $a$-jet and the
$h$-jet. The precision of the mass measurements is therefore 
determined by the precision of muon momentum measurements. We simulate
the toroidal LHC apparatus (ATLAS) detector resolution with ATLFAST~\cite{ATLPHYS}, which smears both the
magnitudes and the directions of the muon momenta, and estimate the mass
measurement error as $\sim M/(100\sqrt{N})$,
where $M$ denotes $M_{a_{\rm dark}}$ or $M_{h_{\rm dark}}$, and $N$ is
the number of corresponding lepton jets. This precision is much better
than the other measurements, as discussed below.  

\begin{figure}[ht!]
\centerline{ \hspace*{0.0cm}
\includegraphics[width=0.48\textwidth]{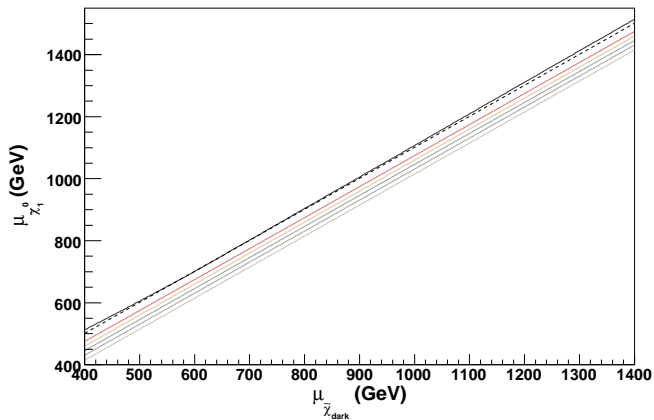}
}
\caption{Contour plot of the number of events consistent with the mass pair
  $(\mu_{\cndark}, \,\mu_{\cnone})$. Starting from the top contour, the
  number of events descends by 1000 from the adjacent contour. Above the
  top contour, the masses are consistent with all 5941  events. The
  dashed straight line has fixed mass difference, which is tangent to
  the top contour. 
\label{fig:mt2}
}
\end{figure}

{\it Measuring $m_{\cnone}$ and $m_{\cndark}$.}  To determine the
masses of $\cnone$ and $\cndark$, we adopt a method 
based on the variable $m_{T2}$ \cite{mt2} on events with 2 $h$- or
$a$-jets. This is an inclusive method since 
that $\cnone$ decays to $\cndark$ in both decay chains in every event.
Alternatively, one could consider more specific processes and 
methods using more kinematic information \cite{kinematics}. For 10
fb${}^{-1}$, we obtain 5941 events with 2 $h/a$-jets after
the cuts in Table \ref{tab:cuts}. We use $\mu_{\cndark}$ and $\mu_{\cnone}$ to denote some trial
masses for $\cnone$ and $\cndark$ and examine the number of events that
are kinematically consistent with each ($\mu_{\cndark}$\,
$\mu_{\cnone}$). As discussed in \cite{min_kinematic},
$(\mu_{\cndark}, \,\mu_{\cnone})$ is consistent with a given event if and only if
$\mu_{\cnone}>m_{T2}(\mu_{\cndark})$, where $m_{T2}$ is calculated
using $\slashchar p_T$ and the momenta of the two lepton jets. Using this fact, we 
easily assign the number of consistent events for each point on the
$(\mu_{\cndark}, \,\mu_{\cnone})$ mass plane (Fig.~\ref{fig:mt2}).

\begin{figure}[ht!]
\centerline{ \hspace*{0.0cm}
\includegraphics[width=0.40\textwidth]{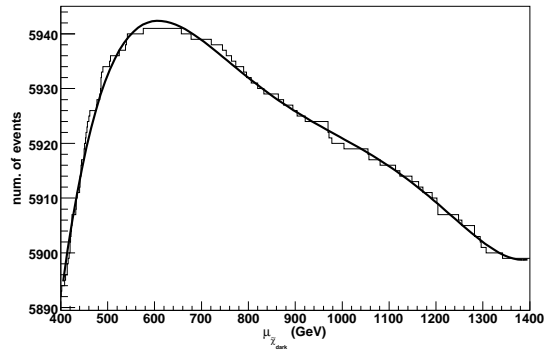}
}
\caption{Number of consistent events along the line with fixed mass
  difference and tangent to the $m_{T2}^{\max}$ contour. 
\label{fig:nevents}
}
\end{figure}

The top contour in Fig.~\ref{fig:mt2} corresponds to the $m_{T2}$
endpoints, $m_{T2}^{\max}(\mu_{\cndark})$.  We note that the
$m_{T2}^{\max}$ contour ``curls up'' on 
both ends with respect to a straight line with constant
mass difference and tangent to the contour. This is caused by the presence
of upstream transverse momentum provided by the squark or gluino decay
as well as the initial state radiation~\cite{UTM}. We count the number
of consistent events along this line, which  is maximized around
$\mu_{\cndark}=m_{\cndark}$ 
(Fig.~\ref{fig:nevents}). We fit the number of events distribution
with a 5th order polynomial and take the maximum position as our
estimation of the DM mass. The error of this measurement is estimated by
repeating the procedure for 10 different datasets and computing the deviation from the 
average value, which gives us $m_{\cndark}=613\pm 12\gev$ (the input
value is 600~GeV) and $m_{\cnone}-m_{\cndark}=101.9\pm 0.8\gev$ (the
input value is 100~GeV)
%~\footnote{A kink structure in the MT2 end point
 % curve can also be used to determine the dark matter
 % mass~\cite{mt2kink2}, although it is difficult to be identified from
 % Fig.~\ref{fig:mt2}. The procedure used in this paper can determine
%  the dark matter mass more precisely.}.
%

{\it Measuring the coupling constant}. As mentioned, we are
interested in the ratio of effective cross sections of $2\,h_\dark$
and $2\,h_\dark \,1\,a_\dark$ events.  Taking the
combinatorial factor into account, this ratio is equal to twice of the
ratio  $R={\rm BR}(\cnone \rightarrow \cndark\,h_{\rm
  dark}\,a_{\rm dark})/ {\rm BR}(\cnone
\rightarrow \cndark\,h_{\rm dark})$. For a fixed ratio of
$M_{h_{\rm dark}}/M_{a_{\rm dark}}=3$, $M_{a_{\rm dark}}\ll m_{\cnone}
- m_{\cndark}$ and $m_{\cnone} - m_{\cndark}\ll
m_{\cnone}$, we obtain an approximate formula
\bea
R\approx\frac{11g^2}{120 \pi^2}\left[
x^2-(8\log{2}+4)x+4(3\log^2{2}+4\log{2}+2)
\right]\,,
\label{eq:ratio2}
\eea
with $x\equiv 2\log{[(m^2_{\cnone} - m^2_{\cndark})/(M_{a_{\rm dark}}\,m_{\cnone})]}$. 
\begin{figure}[ht!]
\centerline{ \hspace*{0.0cm}
\includegraphics[width=0.40\textwidth]{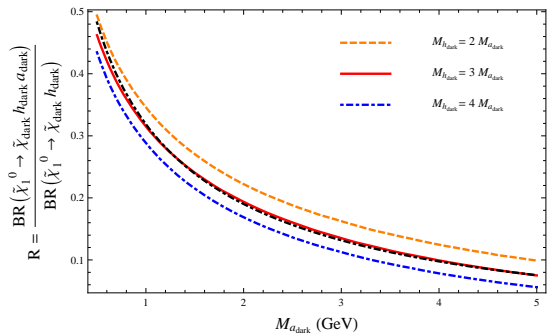}
} 
\caption{The ratio $R$ as a
  function of $M_{a_{\rm dark}}$ for different values of $M_{h_{\rm
      dark}}/M_{a_{\rm dark}}$ and fixed values $m_{\cnone}=700$~GeV
  and $m_{\cndark}=600$~GeV. The ratios are proportional to the gauge
  coupling square, which is chosen to be $g=1$ here. The 
  black dashed line is calculated using the approximate formula in
  Eq.~(\ref{eq:ratio2}). 
}
\label{fig:ratio1}
\end{figure}
Note when $M_{a_{\rm dark}}$ is significantly below 1~GeV, this ratio
$R$ becomes very large and one should include higher order processes
with more $a_{\rm dark}$ radiated.  We also numerically show the ratio
$R$ as a function of $M_{a_{\rm 
dark}}$ for a few different values of $M_{h_{\rm dark}}/M_{a_{\rm
    dark}}$ in Fig.~\ref{fig:ratio1}. 
As can be seen from Fig.~\ref{fig:ratio1}, the approximate formula in
Eq.~(\ref{eq:ratio2}) fits the numerical results very well, and will
be used for error analysis of the gauge coupling measurement. The
double logarithm dependence on $r_a$ or $M_{a_{\rm dark}}$ is due to
the fact that the radiated $a_{\rm 
  dark}$ tends to be collimated with $h_{\rm dark}$.

In our model, the annihilation rate for DM to two $a_\dark$'s or 1
$a_\dark$ plus 1 $h_\dark$ is given by the S-wave result (for a heavy
dark gaugino) 
\bea
\sigma v 
%&=& \frac{g^4}{16\,\pi\,m^2_{\chi_{\rm dark}}} \nonumber \\
=\left(\frac{g}{0.40}\right)^4\left(\frac{m_{\cndark}} {600\,{\rm
      GeV}}\right)^{-2}\times\,2.31\times 10^{-26}{\rm 
  cm}^3/s\,.  
\label{eq:relic}
\eea
where the required rate $2.31\pm 0.07\times 10^{-26}{\rm cm}^3/s$ is
inferred \cite{Bai:2008cf} from the WMAP result $\Omega_{dm}\,h^2=0.113\pm 0.0034$
\cite{WMAP}. An 8\% increase in the annihilation cross section from
the Sommerfeld enhancement is taken into account in
Eq.~(\ref{eq:relic}).  Hence, we have chosen $g=0.40$ and
$m_{\cndark}=600\gev$ to 
give the correct DM relic density. Then $R \approx 0.052$ for
$M_{a_{\rm dark}}=1$~GeV.  
As stated before, this ratio can be measured by counting the number of
events with 2 $h$-jets and the number of events with 2 $h$-jets plus 1
$a$-jet. The latter number is much smaller, which dominates the
error for the $R$ measurement. For $10$ fb$^{-1}$, we expect 220 $2h1a$ 
events before any cuts. However, the $a$-jets tend to be collinear with the
$h$-jets and/or contain soft muons that are not registered by the
detector. This drastically reduces the efficiency for identifying $2h1a$ events to
about 14\% \footnote{The efficiency in general is correlated with the
mass spectrum. However, given the precision of the mass
measurements, this correlation is negligibly small.}, which has to be
taken into consideration when calculating 
the ratio $R$. The number of events is 
reduced to 31 after lepton-jet reconstruction, which results in an
error about 18\% for the measurement of the ratio $R$. Alternatively,
we can treat a lepton jet with 5 or 6 muons as collinear
$h_\dark+a_\dark$ from $\cnone$ three-body decay, which then increases the
number of events to 70 and reduces the error to 12\%.

Having measured the masses and the ratio $R$, we determine the
dark gauge coupling measurement from Eq.~(\ref{eq:ratio2}) as
$g=0.41\pm 0.03$. The error for $g$ is obtained by summing in quadrature the errors of 
all parameters it depends on. Since $M_{a_{\rm dark}}$,
$M_{h_{\rm dark}}$ and  
$m_{\cnone}-m_{{\cndark}}$ can be measured relatively more precisely, the
error is dominated by the errors on $R$ and $m_{\cndark}$. Using
Eq.~(\ref{eq:relic}), we conclude that from the LHC measurements with
10 fb${}^{-1}$, the dark matter relic abundance
can be calculated to be $\Omega^{collider}_{dm}h^2=0.119\pm
0.033$. The result is not as precise as the measurement from WMAP, but
it is certainly very important whether they are consistent with each other.

{\bf Discussions and conclusion.} For illustration, we have assumed that the light mediator only decays to muons. The mediator could decay to electrons as well resulting in lepton jets containing electrons. To identify such lepton jets, we
need to consider electrons isolated from hadronic
activities. Requiring all leptons to be isolated from hadronic jets by
 $\Delta R > 0.4$ and taking the electron acceptance into account
reduce the efficiency for identifying the signal events to around
34\%. The error for the coupling measurement is then 
increased by a factor of three. Moreover, if the light mediator 
only decays to electrons, we will have to determine the mediator's
mass from individual electrons' momenta in an ``electron-jet''. For this
purpose, relatively soft electrons ($\lesssim 10\gev$) are favored
because they can be 
sufficiently separated by the magnetic field before they hit the
electromagnetic calorimeter. Moreover, low momentum electrons are also
well measured with the tracker.

In conclusion, we have described how to test a new class of DM
models with a long-range dark force at the LHC. The light
mediator decays in the detector to a lepton jet, which 
allows us to measure its mass. The DM mass is determined
using the $m_{T2}$ technique. Moreover, we note that when the DM particle is produced, it has a significant rate to radiate an extra light 
mediator. Therefore, we can extract the coupling constant by measuring
the radiation rate. This technique can be generalized to other models
with a dark force. For example, the decay chain in the dark sector may be
longer than the one we have considered, involving more
particles charged under the dark force. In this case, in order to
extract the coupling constant, one needs to carefully reconstruct the
decay chain and include all contributions to the dark radiation. This
and other variations guarantee further studies.

\acknowledgments 
%\vspace*{-.1in}
We thank P. Fox, K.C. Kong and J. Lykken for interesting
discussions. Z.H. is supported in part by the United States Department
of Energy grand ~DE-FG03-91ER40674. Fermilab is operated by Fermi
Research Alliance, LLC under contract ~DE-AC02-07CH11359 with the
United States Department of Energy.

\end{document}